# Gluconate and hexitols effects on C-S-H solubility

Lina Bouzouaid[1], Barbara Lothenbach[2], Alejandro Fernandez-Martinez[3], Christophe Labbez[1]

[1] ICB, UMR 6303 CNRS, Univ. Bourgogne Franche-Comté, FR-21000 Dijon, France

[2] Empa, Concrete & Asphalt Laboratory, Duebendorf, Switzerland

[3] Univ. Grenoble Alpes, Univ. Savoie Mont Blanc, CNRS, IRD, IFSTTAR, ISTerre, 38000 Grenoble, France.

**Abstract:**

This study investigates the effect of gluconate, a carboxylate ion, and three uncharged hexitols, D-sorbitol, D-mannitol and D-galactitol on the solubility of C-S-H. Thermodynamic modeling is used to determine the kind and amount of Ca-organic-silicate-OH complexes that potentially form in the conditions studied. All the organics form complexes with calcium and hydroxide, In addition, heteropolynuclear organics complexes with calcium, hydroxide and silicate are observed at high pH values and high calcium concentrations: $Ca_2Hex_2(H_3SiO_4)_2(OH)_2^0$, $Ca_2Hex_2(H_2SiO_4)(OH)_2^{-2}$ and $Ca_3Gluc_2(H_3SiO_4)_2(OH)_2^0$, with the exception of mannitol. The strength of complexation with silicate decreases in the order gluconate > sorbitol > galactitol. The adsorption of the selected organics on portlandite and C-S-H systems follows the order gluconate >> sorbitol > mannitol ~ galactitol. For C-S-H, a typical Langmuir isotherm is found only when buffered with $Ca(OH)_2$. The adsorption on C-S-H increases with the Ca/Si ratio.

## 1. Introduction

Hydration of cement, involving mainly the anhydrous phases $Ca_3SiO_5$ ($C_3S$) and $Ca_2SiO_4$ (belite), is a complex reaction where two main hydrates are formed: calcium silicate hydrates (C-S-H), and portlandite, $Ca(OH)_2$ (1). It involves the concomitant dissolution of the anhydrate phases and precipitation, from the solution, of the hydrate phases. The passage from solution to these hydrates happens because the ion activity product of the solution (*IAP*) rapidly exceeds the solubility product ($K_{sp}$) of the newly formed solid phases (2). As the concentration and activity of ions build up, the driving force for the nucleation of hydrates ($\Delta\mu$) increases as does the rate of nucleation, *J*. The latter is proportional to $J \propto \exp\left(-\Delta\mu^{-2} v_m^2 \gamma^3 / kT\right)$ with $k_B$ being the Boltzmann constant and *T* the temperature and where $v_m$ and $\gamma$ are the molecular volume and surface tension of the hydrate. The driving force is defined by the change in chemical potential ($\Delta\mu$) of the crystallizing ions, and measures the free energy response to transferring the ions from the solution to the solid, $\Delta\mu = k_B T (\ln IAP - \ln K_{sp})$. The same reasoning applies to the dissolution of anhydrous phases in a symmetric way. The activity of the dissolving and crystallizing ions is an essential thermodynamic quantity if one wants to quantify and understand the hydration of cement. However, ion activities are strongly influenced by complex formation, e.g between calcium and organics. In the present study we aim at extending our previous work (3) on the calcium activity in the presence of four cement organic retarders, namely sodium D-gluconate and three hexitols (D-mannitol, D-sorbitol and D-galactitol), to the speciation and activity of calcium and silicate solution species in the presence of the same organics.

Superplasticizers, in particular, and organic molecules, in general, often behave as retarders when used in concrete (4)(5)(6). At low percentage, less than 1% wt of Portland cement, they retard the initial and final setting times of cement (7)(8). Several studies have been conducted to highlight their influence on the hydration of cements, in order to obtain a better understanding of the mechanisms of retardation and the setting times, but still the physical and chemical mechanisms responsible for this retardation remained elusive (9)(10)(11)(12)(13). Using small organic molecules as a model for the more complex superplasticizers, previous works have been directed to decipher which step ($C_3S$ dissolution or C-S-H nucleation/growth) of the cement hydration is specifically impacted by organics. These includes the charged molecule gluconate known to behave as a strong retarder and hexitols (14)(15)(16)(17)(18). For some hexitols and gluconate, the dissolution rate of $Ca_3SiO_5$ was observed to be only marginally affected (19) and they were thus conjecture to act mainly as inhibitors of the nucleation/growth of C-S-H (20) and portlandite (21). The neutral hexitol molecules were observed to retard the hydration of cement much less than the charged gluconate molecule, and their retardation power was found to depend on their stereochemistry (19) (20) (22).

In a recent study, we investigated the effect of gluconate, D-sorbitol, D-galactitol and D-mannitol on calcium speciation at high pH values (3). It was found that the sorption of organics on portlandite as well as the strength of organics complex formation with Ca decreases in the order gluconate >> sorbitol > mannitol ~ galactitol, similar to the organics tendency to retard the $C_3S$ hydration (19): gluconate >> sorbitol > galactitol > mannitol. In the presence of portlandite, polynuclear complexes with calcium and hydroxide such as $Ca_3Gluc_2(OH)_4^0$ dominate the Ca-speciation with gluconate, while ternary $CaHexOH^+$ and polynuclear $Ca_2Hex_2OH_4^0$ complexes are dominant in the presence of the hexitols (3).

In the present paper, we investigate the combined speciation of calcium, silicate and organics (gluconate, D-sorbitol, D-galactitol or D-mannitol) at concentrations and pH values relevant for cementitious systems. Complex formation is experimentally studied based on solubility measurements of calcium silicate hydrate (C-S-H) at various Ca/Si, ranging from 0.75 to 1.5, in the presence of increasing amount of organics. The strength and the various types of calcium and silicate complexes with the organic molecules is determined with a speciation model.

## 2. Materials and methods

### 2.1 Materials

The sorption and complex formation of organics were studied on pre-synthesized C-S-H. A large stock of C-S-H was synthesized with a Ca/Si of 0.75 by adding 4.2 g of lime and 5.9 g of silicate. The reactants were mixed with boiled and degassed water, within 250 mL PE containers protected from air, using a liquid to solid ratio (L/S) of 25 to obtain a homogeneous suspension. The containers were placed on a shaking table with a stirring speed of 180 rpm for four weeks at T=23°C. Before filtration, the PE containers were left to stand for 24 hours allowing the solid to settle. The supernatant was filtered through a 40 μm sintered glass filter. The solid phase was dried under suction in a desiccator in presence of silica gel at room temperature for two weeks and then stored in desiccator under vacuum in the presence of silica gel. It is important to mention here that C-S-H with any C/S can rapidly and easily be obtained from the pre-synthesized powder of C-S-H with C/S = 0.75 by equilibrating it in an aqueous solution with the appropriate concentration of $Ca(OH)_2$, see the solubility diagram of C-S-H in e.g. ref (23)

The lime powder used was obtained by calcining calcium carbonate $CaCO_3$ (WR Analar Normapur, 98.5-100%) at 1000°C for 24 hours. Aerosil 200 silica (Evonik industries) was used, a hydrophilic fumed silica with a specific surface area of 200 m$^2$/g and a high reactivity. Different organic compounds were used for the solubility experiments: potassium gluconate ($C_6H_{11}KO_7$, Sigma-Aldrich, ≥97% purity), D-sorbitol ($C_6H_{14}O_6$, Sigma-Aldrich, ≥99% purity), D-mannitol (C$_6$H$_{14}$O$_6$, Sigma-Aldrich, ≥99% purity), and D-galactitol (C$_6$H$_{14}$O$_6$, Sigma-Aldrich, ≥99% purity). Note that the three hexitols have the same structural units but have a different configuration as illustrated in Figure *1*.

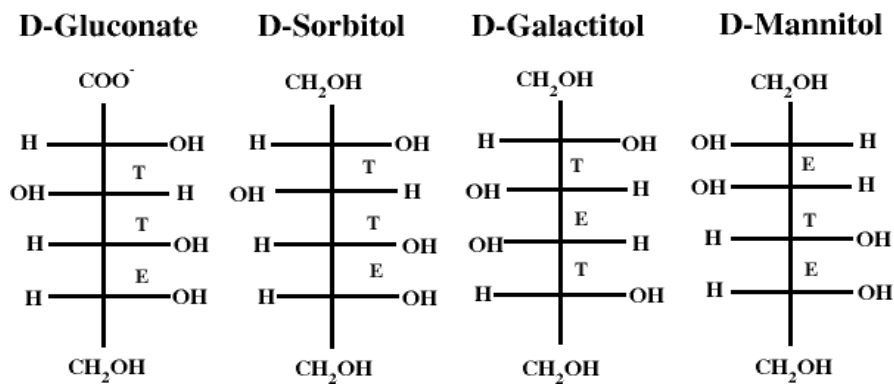

Figure 1: Structure of (from left to right) D-gluconate, D-sorbitol, D-galactitol and D-mannitol. "T" corresponds to the threo diastereoisomer configuration. "E" corresponds to erythro diastereoisomer configuration. Configuration refers here to that of two adjacent stereo-centers (carbons) relative to each others.

*2.2     Solubility experiments*

Four different series of samples were prepared in a glove box for each organic molecule: C-S-H equilibrated with 0 mM, 11 mM and 20 mM calcium, added in the form of CaO, as well as a series with saturated lime solution. For each experiment 400 mg of C-S-H (C/S =0.75) were used as a solid buffer, enclosed in a dialysis membrane and placed in a 250 mL polypropylene flask filled with 250 mL of CO$_2$-free solution with different amounts and type of organic molecule (see Figure 2). Note that in the case of gluconate only 200 mg of C-S-H was used. Based on mass balance calculations (23) the final Ca/Si in C-S-H corresponded to 0.75 (0 mM Ca), 1.1 (11 mM Ca), 1.3 (20 mM Ca), and to 1.5 (saturated lime solution; initial Ca/Si in solid = 0.75).

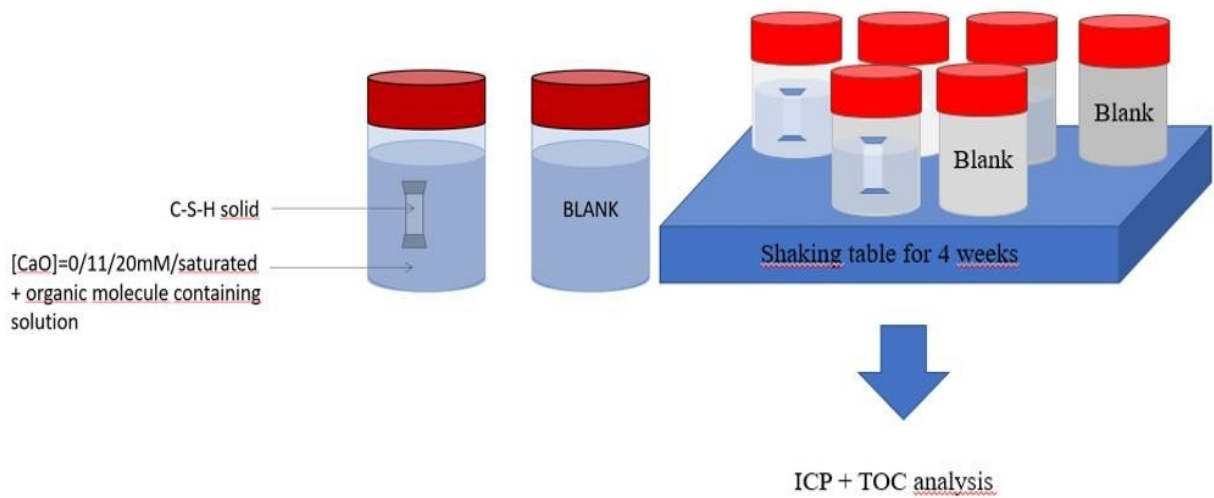

Figure 2: Schematic representation of the different series of C-S-H with 0 mM, 11 mM, 20 mM and saturated with Ca(OH)$_2$, used for the solubility and adsorption experiments. The flask used for the solubility measurement contains a dialysis bag filled with 200mg (Gluconate) or 400 mg (Hexitols) of C-S-H powder, immersed in 250 mL of solution containing the organic molecule. The second flask containing the organic molecule solution only is used as a blank/reference to verify the organic concentration introduced initially. This allows the determination of the organic adsorption on C-S-H by mass balance based on the measured difference. Note that the same stock solution is used for each sample pair.

The dialysis membranes (Spectra / Por, MWCO 12-14 kD) were dipped before use in distilled water for 30 minutes to remove any organic residues and dried in a desiccator overnight. The dialysis bags were closed with polyamide clamps (Carl Roth, length 50 mm). Finally, the samples were stored in plastic barrels filled with N$_2$ gas to guarantee CO$_2$ free conditions, and placed on a shaking table during four weeks at 23°C to equilibrate. In each case, in addition to the samples containing the dialysis bag filled with C-S-H, blank samples containing only the lime and organic molecule solution were prepared and used as reference. The pH values and total elemental concentrations were measured in the solution after removing the dialysis bags. The pH electrode was calibrated using Sigma Aldrich buffer (pH 4, 7, 9 and 12). It should also be mentioned that the pH was not used as a fitting parameter for the complex formation. The reason being the difficulty to get accurate and reliable pH values inherent to the impossibility of measuring the proton activity alone, see e.g. ref (34). The total concentration of the elements Ca and Si were measured by inductively coupled plasma-optic emission spectroscopy (ICP-OES 5110, Agilent) in diluted solution acidified with HNO$_3$. Five or six different organic concentrations were studied, with a starting concentrations between 0 mM to 200 mM (detailed values are given in the Supplementary Information). The bulk concentration of organics at equilibrium was measured as total organic content with a TOC VCPN instrument (Shimadzu). This tool is based on the oxidation of organic molecules contained in solution by gaseous oxygen with a platinum-based catalyst in an oven raised to a temperature of 720°C. The $CO_2$ formed is detected by Infra-Red Non-Dispersive (NDIR). The detection threshold for this device is very low (4µg/L).

*2.3 Thermodynamic simulation and complexation constants*

The solubility of C-S-H and the activity of calcium, silicate and hydroxide solution species at equilibrium were fitted using a speciation model solved by the geochemical software PHREEQC version 3 (3.6.2-15100) and the WATEQ4f database. The activity of the ion species $i$, $a_i$, was calculated according to $a_i = \gamma_i \cdot m_i$, where $\gamma_i$ is the activity coefficient and $m_i$ the molality in mol/kg $H_2O$. The activity coefficients were calculated in PHREEQC for major species with the WATEQ Debye Hückel equation:

$$\log \gamma_i = \frac{-A z_i^2 \sqrt{I}}{1+B d_i \sqrt{I}} + b_i I \qquad (1)$$

where $z_i$ denotes the charge of species $i$, $I$ the molal ionic strength, $d_i$ the individual ion-size parameter, and $b_i$ is an ion specific parameter. $A$ and $B$ are pressure- and temperature-dependent coefficients. Equation 1 is applicable up to approximately 1 M of ionic strength. For non tabulated species, the Davies equation is used to calculate the activity coefficients in PHREEQC.

**Table 1.** Solubility and complex formation constants of calcium, silicate, hydroxide and gluconate ($Gluc^-$) containing species at standard conditions (1 bar, 25°C) and I = 0 reported in literature and determined in the present study.

|  | $\log_{10} K^a$ | References |
|---|---|---|
| **Solids** | | |
| $Ca(OH)_2 + 2H^+ = Ca^{+2} + 2H_2O$ | 22.80 | 24 |
| $SiO_2(amorphous) + 2H_2O = H_4SiO_4^0$ | -2.71 | 23 |
| **Aqueous species** | | |
| $Ca^{+2} + OH^- = CaOH^+$ | $1.22^b$ | 24 |
| $H_4SiO_4^0 = H_3SiO_4^- + H^+$ | -9.83 | 25 |
| $H_4SiO_4^0 = H_2SiO_4^{-2} + 2H^+$ | -23.10 | 25 |
| $Gluc^- + H^+ = GlucH^0$ | 3.64 | $26^c$ |
| $Gluc^- + OH^- = GlucOH^{-2}$ | -0.44 | $27^c$ |
| $Ca^{+2} + Gluc^- = CaGluc^+$ | 1.56 | $28^c$ |
| $Ca^{+2} + 2Gluc^- = CaGluc_2^0$ | 2.85 | 3 |
| $Ca^{+2} + OH^- + Gluc^- = CaGlucOH^0$ | 3.95 | 3 |
| $2Ca^{+2} + 4 OH^- + 2Gluc^- = Ca_2Gluc_2(OH)_4^{-2}$ | 11.25 | 3 |
| $3Ca^{+2} + 4 OH^- + 2Gluc^- = Ca_3Gluc_2(OH)_4^0$ | 16.10 | 3 |
| $3Ca^{+2} + 2 OH^- + 2Gluc^- + 2H_3SiO_4^- = Ca_3Gluc_2(H_3SiO_4)_2(OH)_2^0$ | 20.45 | This study |

[a] The standard error on the fitted Log K values is estimated to be less than 0.05. [b] The potential formation of a $Ca(OH)_2^0$ complex with a $\log_{10} K = 1.85$ for the reaction $Ca^{2+} + 2OH^- = Ca(OH)_2^0$ (29) was initially considered, but its presence or absence had no significant effect at pH values below 13, such that it was neglected in the final calculations. [c] Values given in the present table are those reported by the cited references extrapolated to 0 M ionic strength using the WATEQ Debye Hückel equation, see (3) for more details.

**Table 2.** Complex formation constants of calcium, silicate, hydroxide and sorbitol (Sorb), mannitol (Man) and galactitol (Gal) containing species at standard conditions (1 bar, 25°C) and I = 0 reported in literature and determined in the present study

|  | $\log_{10} K^a$ | References |
|---|---|---|
| $Ca^{+2} + Sorb^0 = CaSorb^{+2}$ | 0.10 | 3 |
| $Ca^{+2} + Sorb^0 + OH^- = CaSorbOH^+$ | 2.85 | 3 |
| $2Ca^{+2} + 2Sorb^0 + 4OH^- = Ca_2Sorb_2(OH)_4^0$ | 9.75 | 3 |
| $2Ca^{+2} + 2Sorb^0 + 2OH^- + 2H_3SiO_4^- = Ca_2Sorb_2(H_3SiO_4)_2(OH)_2^0$ | 16.00 | This study |
| $2Ca^{+2} + 2Sorb^0 + 4OH^- + H_2SiO_4^{-2} = Ca_2Sorb_2(H_2SiO_4)(OH)_4^{-2}$ | 13.60 | This study |
| $Ca^{+2} + Man^0 = CaMan^{+2}$ | -0.36 | 3 |
| $Ca^{+2} + Man^0 + OH^- = CaManOH^+$ | 2.65 | 3 |
| $2Ca^{+2} + 2Man^0 + 4OH^- = Ca_2Man_2(OH)_4^0$ | 9.65 | 3 |
| $2Ca^{+2} + 2Man^0 + 2OH^- + 2H_3SiO_4^- = Ca_2Man_2(H_3SiO_4)_2(OH)_2^0$ | NA* | This study |
| $2Ca^{+2} + 2Man^0 + 4OH^- + H_2SiO_4^{-2} = Ca_2Man_2(H_2SiO_4)(OH)_4^{-2}$ | NA* | This study |
| $Ca^{+2} + Gal^0 = CaGal^{+2}$ | -0.53 | 3 |
| $Ca^{+2} + Gal^0 + OH^- = CaGalOH^+$ | 2.80 | 3 |
| $2Ca^{+2} + 2Gal^0 + 4OH^- = Ca2Gal_2(OH)_4^0$ | 9.29 | 3 |
| $2Ca^{+2} + 2Gal^0 + 2OH^- + 2H_3SiO_4^- = Ca_2Gal_2(H_3SiO_4)_2(OH)_2^0$ | 14.60 | This study |
| $2Ca^{+2} + 2Gal^0 + 4OH^- + H_2SiO_4^{-2} = Ca_2Gal_2(H_2SiO_4)(OH)_4^{-2}$ | 13.00 | This study |

[a] The standard error on the fitted $\log_{10} K$ values is estimated to be less than 0.05. *NA: Not available: no constants could be fitted in this study due to the weak complexation of mannitol with Si.

**Table 3.** Solubility and surface complexation constants of the C-S-H phases at standard conditions (1 bar, 25°C) and I = 0 (23)

|  | $\log_{10} K$ |
|---|---|
| α-C-S-H: $Ca_4Si_5O_{16}H_4^0 + 8H^+ + 4H_2O = 4Ca^{+2} + 5H_4SiO_4^0$ | 53.5 |
| β-C-S-H: $Ca_2Si_2O_7H_2^0 + 4H^+ + H_2O = 2Ca^{+2} + 2H_4SiO_4^0$ | 29.6 |
| γ-C-S-H: $Ca_6Si_4O_{15}H_2^0 + 12H^+ + H_2O = 6Ca^{+2} + 4H_4SiO_4^0$ | 104.5 |
| $\equiv SiOH^0 = \equiv SiO^- + H^+$ | -9.8 |
| $\equiv SiOH^0 + Ca^{+2} = \equiv SiOCa^+ + H^+$ | -7.0 |
| $\equiv SiOH^0 + Ca^{+2} + OH^- = \equiv SiOCaOH^0$ | -9.0 |
| $2\equiv SiOH^0 + H_4SiO_4^0 = (\equiv SiO)_2Si(OH)_2^0 + 2H_2O$ | 5.8* |
| $2\equiv SiOH^0 + Ca^{+2} = (\equiv SiO)_2Ca^0 + 2H^+$ | -11.4 |

* The indicated value is for β-C-S-H and γ-C-S-H; log K = 4.4 for α-C-S-H

The fitted complex formation constants between calcium, silicate, hydroxide and the organics used in the present study are listed in Tables 1 and 2. The quaternary complexes were determined by simple and systematic variations of the ternary complexes determined in Kutus and Pallagi's works and further confirmed in ref (3) in the CaO-Org-H$_2$O system. This amounted to keep the Ca/Org stoechiometric ratio constant, substituting OH by silicate monomer or adding silicate mononer and varying the ionization degree of the latter. The calculations made to identify the composition of the solution at equilibrium with C-S-H in suspensions with and without organic molecules used the surface complexation model of Haas and Nonat, which takes into account the variation of the solubility and stoichiometry of C-S-H phases (23). The values of the solubility constants and surface complexation of the phases α, β and γ of C-S-H are listed in Table 3.

## 3. Results and discussion

### 3.1 Solubility of C-S-H in presence of the organics

#### 3.1.1. Hexitols

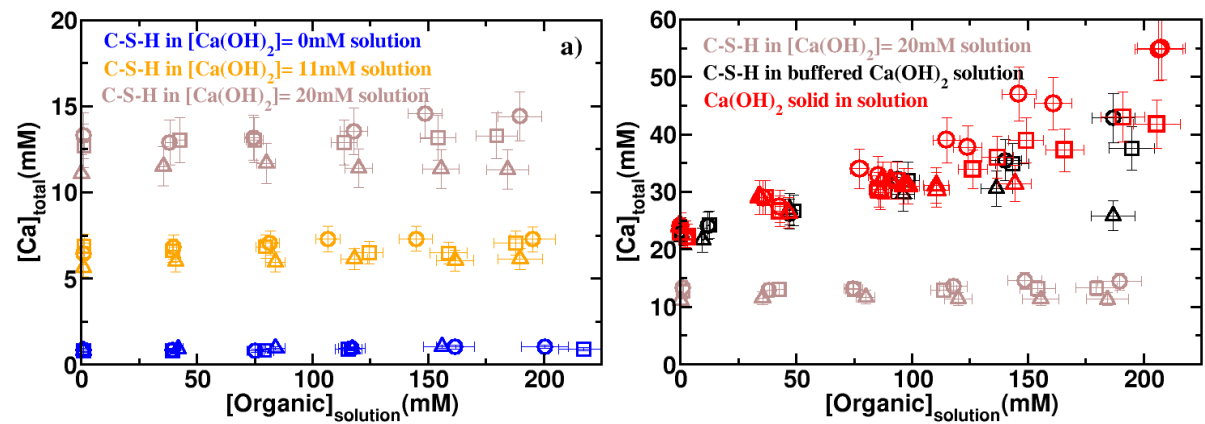

Figure 3: Measured solution concentrations of calcium in equilibrium with C-S-H and portlandite (red symbols, from (3)) as a function of the equilibrium concentration of hexitols. Sorbitol, mannitol and galactitol are represented by circles, squares and triangles, respectively. The C-S-H (C/S = 0.75) containing dialysis bags were immersed in 250 mL of ultra-pure and degassed water containing initially 0 mM Ca(OH)$_2$ (blue symbols), 11 mM Ca(OH)$_2$ (orange symbols), and 20 mM Ca(OH)$_2$ (grey symbols). The corresponding initial pH are 10.6, 12.1, 12.4, see the sup. info. The black symbols show the results for C-S-H in buffered portlandite conditions[1], pH 12.7, see also Figure S.1.

The evolution of the equilibrium calcium concentration in solution resulting from the solubility of C-S-H upon increasing the hexitol concentration is shown in Figure 3. The obtained results are further compared to those obtained in our previous study with portlandite (3), see Figure 3-b. In the presence of increasing hexitol concentrations the measured total calcium concentration is observed to raise only moderately independently of whether sorbitol, mannitol or galactitol is used. For C-S-H with the lowest C/S studied (0.75), the total equilibrium concentration of calcium remains almost insensitive to the hexitol addition. Only for C-S-H buffered with portlandite[1] (with C/S =1.5 in C-S-H) a significant increase in equilibrium calcium concentrations from 20 mM to approximately 50 mM Ca can be observed, see Figure 3-b, indicating complexation between hexitol and calcium in solution. We can further note from Figure 3-b that the equilibrium calcium concentrations obtained from the solubility experiments with C-S-H in a buffered Ca(OH)$_2$ solution (C/S 1.5) and in experiments where only portlandite was present (for details see (3)) show the same trends and concentrations, independent whether sorbitol, mannitol or galactitol has been added. This similarity clearly indicates that the raise in equilibrium calcium concentration is to a great extent due to the formation of the ternary Hex$_a$-Ca$_b$-OH$_c$ complexes (Ca$_1$Hex$_1$(OH)$_1^+$, Ca$_2$Hex$_2$(OH)$_4^0$) already identified in our previous study (3). We can expect that the same is true for C-S-H with lower C/S, although the concentrations of Hex$_a$-Ca$_b$-OH$_c$ complexes are lower due to the lower calcium concentration. The impact of the various hexitols on the equilibrium calcium concentration follows the strength of the hexitol complex formation with calcium found with portlandite (3): it increases in the order galactitol ~ mannitol < sorbitol.

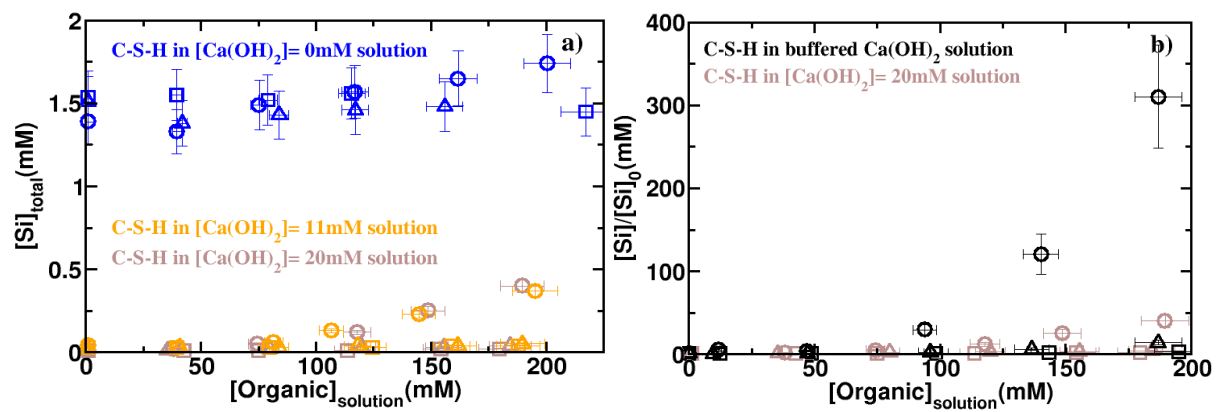

Figure 4: Measured solution concentrations of a) silicate and b) relative increase of the silicate concentration in the solution, in equilibrium with C-S-H prepared in the same conditions as in Figure 3. The relative increase of silicate is expressed as the ratio of the equilibrium silicate concentration to the equilibrium silicate concentration without hexitols, [Si]$_0$. Circle, square and triangle symbols represent sorbitol, mannitol and galactitol, respectively.

The variation of silicate concentration in solutions at equilibrium with C-S-H as a function of the organic concentrations is shown in Figure 4-a. Similarly to calcium, the equilibrium silicate concentration rises with the concentration of the hexitols at all C/S studied. At the lowest C/S ratio studied (0.75, blue symbols, see Figure 4-a), a slight increase of silicate concentration from 1.5 mM in the absence of hexitols to 1.7 mM in the presence of 200 mM

---

1 We here note that buffering solutions with an excess of solid Portlandite allows to maintain constant the mean ionic activity of Ca(OH)$_2$ in the solution: $a_{Ca(OH)_2} = a_{Ca^{2+}} + a_{OH}^2 =$ Cste hexitol is observed. The increase in the case of sorbitol is slightly more pronounced than for mannitol and galactitol. At C/S =1.1 (11 mM Ca(OH)$_2$), the Si concentration remains below 0.05 mM in the presence of mannitol and galactitol, whereas for sorbitol it increases almost ten times up to 0.37 mM of silicate in the presence of 195 mM sorbitol. At higher C/S (1.3 and 1.5), the increase in silicate concentration remains weak with mannitol and galactitol and strong with sorbitol, as illustrated in Figure 4b.

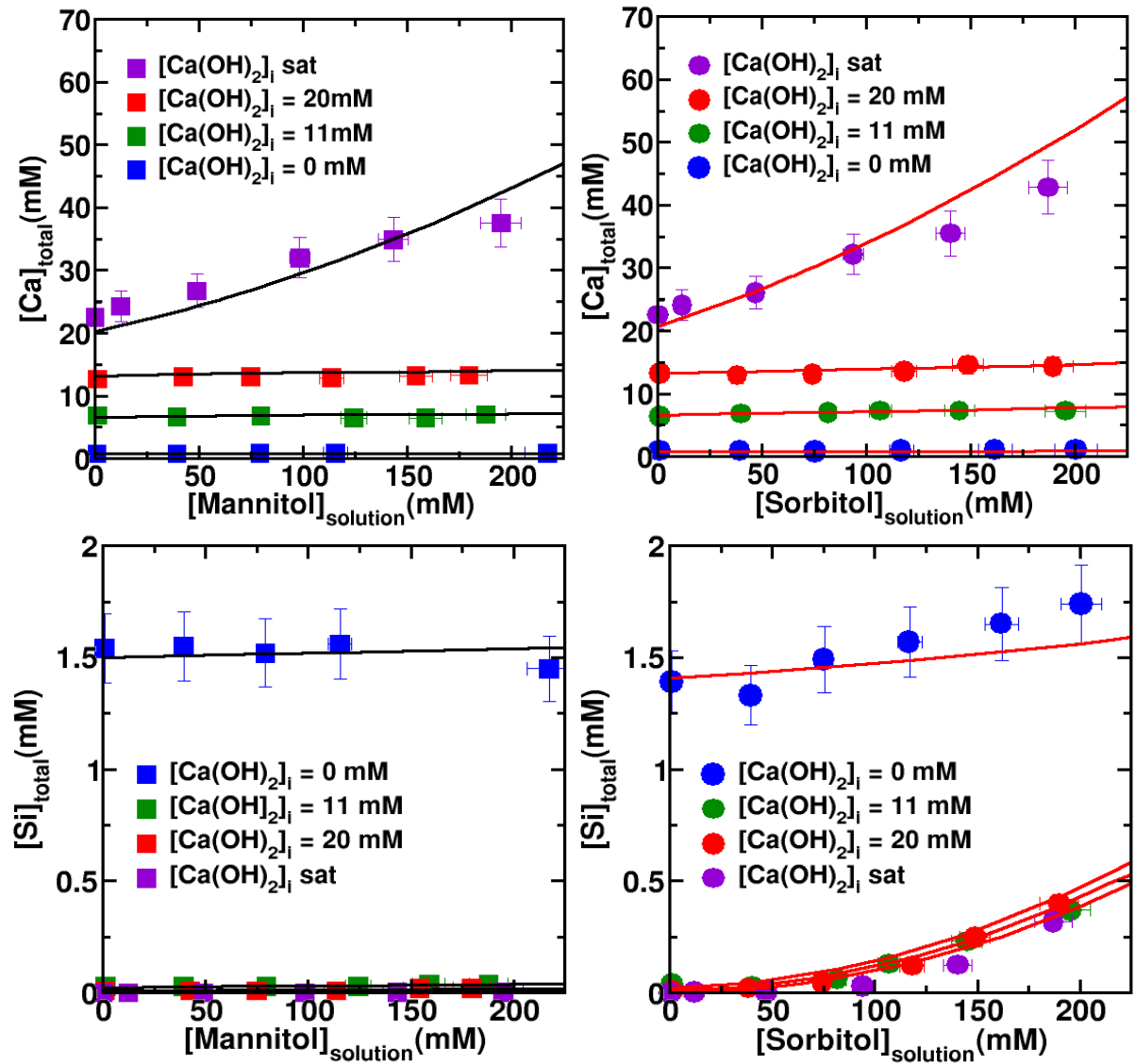

Figure 5: Total experimental (symbols) and simulated (lines) concentrations of aqueous calcium and silicate species at equilibrium with C-S-H immersed in 250 mL of solution containing various initial concentrations of Ca(OH)$_2$, [Ca(OH)$_2$]$_i$, and varying the equilibrium concentration of hexitols.

The raise in silicate and calcium concentration in response to hexitol addition becomes stronger when the C/S of C-S-H is increased, i.e. at high pH and calcium concentrations, while at the lowest C/S studied, C/S = 0.75, the addition of hexitols has no significant impact.

At C/S 1.5 ($Ca(OH)_2$ buffered systems) the silicate concentration increases by more than 2 orders of magnitude in presence of ~200 mM of sorbitol, see Figure 4-b. The impact of the different hexitols on the silicate concentration is clearly different and increases in the following order: mannitol < galactitol << sorbitol, i.e. similar to that observed for calcium. These results indicate the formation of quaternary complexes, involving silicate-calcium-hydroxide and hexitols for galactitol and sorbitol, while none or very weak complexes are formed in the case of mannitol. Below, we shall show that such complexes derived from the ternary complex $Ca_2Hex_2(OH)_4^0$ identified in our previous study can give an accurate description of the solubility data of C-S-H presented above.

The measured and calculated silicate and calcium concentrations are compared in Figure 5. The Ca and Si concentration were calculated using the known Ca-hexitol complexes ($CaHexOH^+$ and $Ca_2Hex_2(OH)_4^0$) as summarized in Table 2 together with two quaternary Si-complexes, $Ca_2Hex_2(H_2SiO_4)(OH)_4^{-2}$ and $Ca_2Hex_2(H_3SiO_4)_2(OH)_2^0$, which were fitted here. The comparison of the calculated with the experimental calcium and silicate concentrations in equilibrium with C-S-H in Figure 5 shows in both cases significant increase of the Ca and silicate concentrations at high C/S and high pH values. The figures for galactitol are presented in the Supplementary Information, see Figure S.2.

As the silicate complexes with organics occur mainly at high Ca-concentrations and pH values, we hypothesized that polynuclear complexes similar to the main Ca-hexitol complexes are present in these conditions, namely $Ca_2Hex_2(OH)_4^0$: a neutral $Ca_2Hex_2(H_3SiO_4)_2(OH)_2^0$ and a negatively charged $Ca_2Hex(H_2SiO_4)(OH)_4^{-2}$ complex. The use of only one of these complexes was not able to reproduce the experimental data for sorbitol and galactitol adequately, nor the use of simpler complexes, i.e. $CaHexH_2SiO_4^0$ or $CaHex(H_3SiO_4)_2^0$. For mannitol no Si- containing complexes were found as detailed in Table 2.

It should also be noted that the modeling of the silicate and calcium concentrations using classical speciation modeling (e.g. PHREEQC or GEMS) is strictly equivalent if one uses complexes with different size (stoichiometric coefficient) but same stoichiometry, i.e. $Ca_yHex_y(H_3SiO_4)_y(OH)_y^0$ and $Ca_zHex_{0.5z}(H_2SiO_4)_{0.5z}(OH)_{2z}^{-z}$ with $y > 2$ and $z > 2$ (provided that the formation constants are modified accordingly) or if one defines a different deprotonation degree (n) for the silicate and hexitols in the complex providing that the overall charge of the complex is maintained constant (i.e. number of $OH^-$ groups adapted accordingly). In other words, the use of $Ca_yHex_y(H_{3-n}SiO_4)_y(OH)_{y-n}^0$ and $Ca_y(Hex^{-n})_y(H_3SiO_4)_y(OH)_{y-n}^0$ as well as $Ca_z(Hex^{-n})_{0.5z}(H_2SiO_4)_{0.5z}(OH)_{2(z-n)}^{-z}$ and $Ca_z(Hex)_{0.5z}(H_{2-n}SiO_4)_{0.5z}(OH)_{2(z-n)}^{-z}$ leads exactly to the same modelling results. Again the choice of y=2 and z=2 was guided by previous works by Pallagi et al. (27) (28) and ourselves (3), but is somewhat arbitrary. Indeed, preliminary speciation simulations in the framework of the primitive model indicate that the complexes tend to form clusters that grow in size (i.e. increase of y and z) with the organic and calcium concentrations. Also, at high calcium and organic concentrations, we experimentally observed that the solutions were extremely difficult to filtrate, which was one of the motivation to separate the solids from the solution with a dialysis membrane. This is another indication of the formation of large complex clusters and of the variation of the size of the complex clusters with the equilibrium conditions.

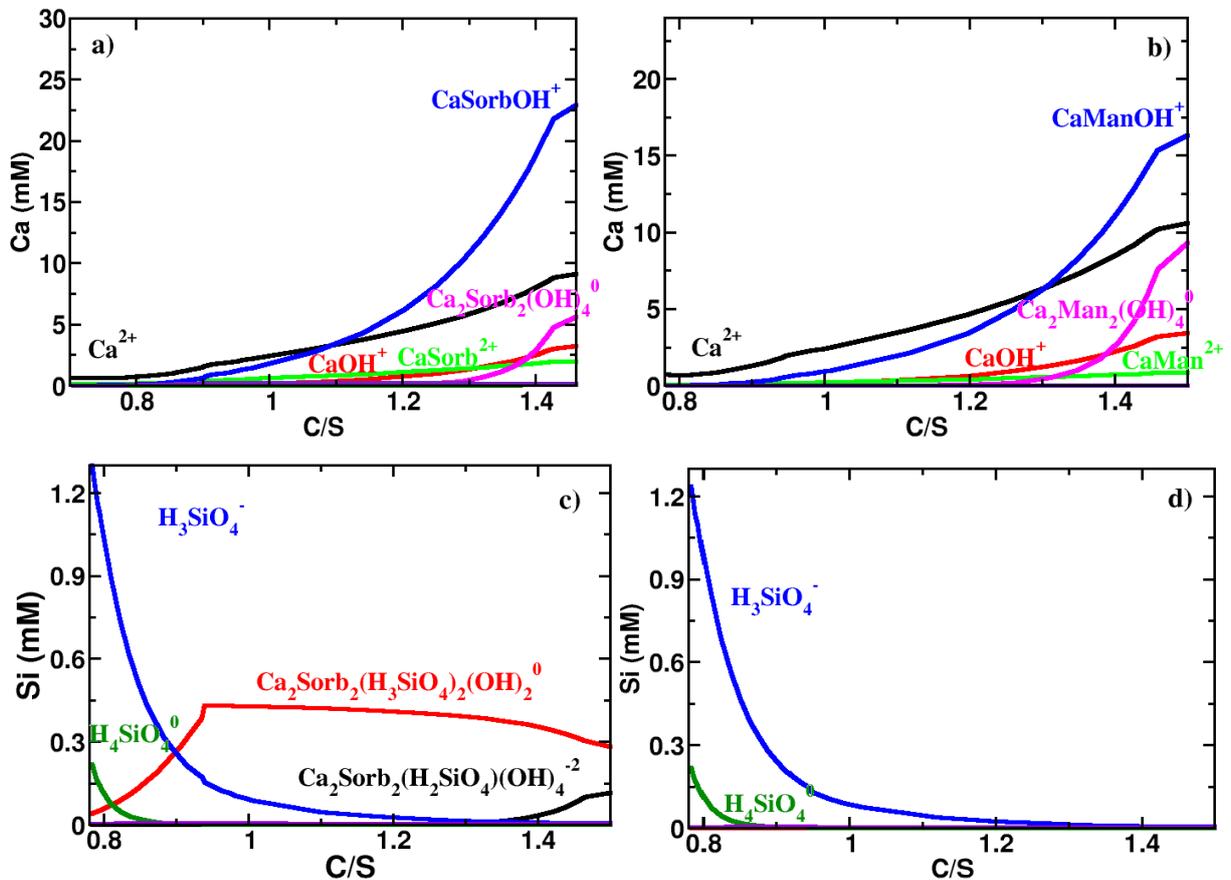

Figure 6: Simulated speciation of calcium and silicate species in a 250 mL solution containing 200 mM of hexitols in equilibrium with 200 mg of C-S-H with varying C/S ratio. (a) and (c) provides the results for sorbitol; (b) and (d) those for mannitol. The distribution diagram can be found in the sup. info., see Figure S3 and S4.

The use of CaSorbOH$^+$, Ca$_2$Sorb(OH)$_4^0$ together with Ca$_2$Sorb$_2$(H$_3$SiO$_4$)$_2$(OH)$_2^0$ and Ca$_2$Sorb$_2$(H$_2$SiO$_4$)(OH)$_4^{-2}$ allows us to reproduce the increase in Ca concentrations at all Ca/Si ratios studied as well as the moderate increase in Si concentrations in the high range of Ca concentrations and pH values. The speciation of Si is dominated by H$_3$SiO$_4^-$ at low Ca(OH)$_2$ concentration (C/S = 0.75) and by Ca$_2$Sorb$_2$(H$_3$SiO$_4$)$_2$(OH)$_2^0$ at C/S of 1 and above, as shown in Figure 6 and Figure S4. We can further note that the negatively charged

$Ca_2Sorb_2(H_2SiO_4)_2(OH)_4^{-2}$ complex becomes only important at high pH values and Ca concentrations, typically in the buffered portlandite systems, where the concentrations of free silicates ($H_3SiO_4^-$, $H_2SiO_4^{-2}$) are negligibly small (more than ten times smaller than the quaternary complexes). On the other hand, the quaternary complexes has a negligible impact on the total calcium concentration as seen in Figure 6-a and Figure S3. To further illustrate the importance of the quaternary complexes on the C-S-H solubility a comparison of the measured and predicted total equilibrium concentrations of Si and Ca when those complexes are ignored is provided in the supplementary materials, see Figure S5.

In the case of mannitol, a very good fit of the experimental data is also obtained, see Figure 5, without the need of the $Ca_2Man_2(H_3SiO_4)_2(OH)_2^0$ and $Ca_2Man_2(H_2SiO_4)(OH)_4^{-2}$ complexes (see Table 2), as no significant increase of Si-concentration was observed. This is shown by the simulated speciation data plotted in Figure 6 where no polynuclear calcium organic silicate complexes are found, contrary to the case of sorbitol and galactitol. The $H_4SiO_4^0$, $H_3SiO_4^-$, and $H_2SiO_4^{-2}$ species thus dominate the silicate speciation at all Ca/Si.

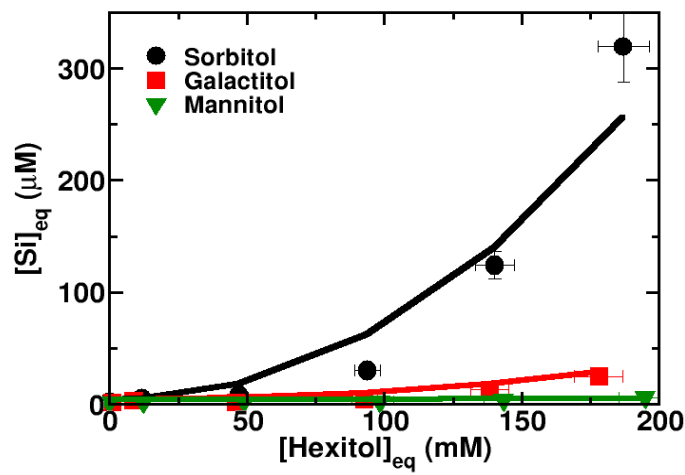

Figure 7: Total experimental (symbols) and simulated (lines) concentrations of silicate species in the aqueous phase in equilibrium with C-S-H buffered with portlandite (C/S 1.5) and varying the equilibrium concentration of hexitols.

Similarly to sorbitol, the use of $CaGalOH^+$, $Ca_2Gal(OH)_4^0$, $Ca_2Gal_2(H_3SiO_4)_2(OH)_2^0$ and $Ca_2Gal_2(H_2SiO_4)(OH)_4^{-2}$ complexes allows us to fit accurately the measured calcium and silicate concentrations in presence of increasing amount of galactitol. Note that the fitted constants for $Ca_2Gal_2(H_3SiO_4)_2(OH)_2^0$ and $Ca_2Gal_2(H_2SiO_4)(OH)_4^{-2}$ as given in Table 2 are several log units weaker than those obtained for $Ca_2Sorb_2(H_3SiO_4)_2(OH)_2^0$ and $Ca_2Sorb_2(H_2SiO_4)(OH)_4^{-2}$ consistent with the much weaker complexation of galactitol with calcium ions (3). This is also illustrated in Figure 7, which compares the solubility of C-S-H buffered with portlandite (C/S 1.5) in the presence of sorbitol, galactitol and mannitol.

Compared to sorbitol, a much lower amount of silicate is found in solution for galactitol but clearly more than in the case of mannitol for which no or only very weak polynuclear silicate complexes are formed. $Ca_2Gal_2(H_3SiO_4)_2(OH)_2^0$ and $Ca_2Gal_2(H_2SiO_4)_2(OH)_4^{-2}$ are predicted to dominate the solution speciation of silicates only at high C/S ratio, comparable to sorbitol.

*3.1.2 Sodium gluconate*

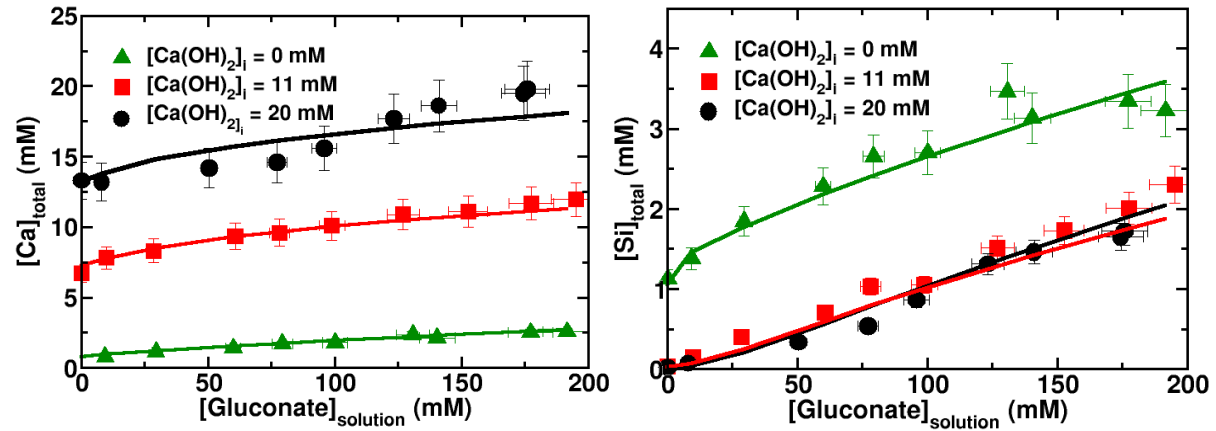

Figure 8: Experimental (symbols) and simulated (lines) concentrations of aqueous calcium and silicate species at equilibrium with C-S-H for various initial concentrations of $Ca(OH)_2$ in the solution, $[Ca(OH)_2]_i$, as detailed in materials and methods. The simulations were performed with the complexes listed in Table 1.

The evolution of calcium and silicate concentration in solution at equilibrium with C-S-H as a function of gluconate concentrations is shown in Figure 8. Compared to the hexitols, the measured total calcium concentration is observed to raise more strongly in the presence of gluconate at all C/S studied, indicating a strong complexation between gluconate and calcium in solution. A four times higher calcium concentration is observed in the presence of 192 mM of gluconate for the lowest C/S studied ($[Ca(OH)_2]_i$ = 0 mM) than in its absence, while at C/S=1.3 ($[Ca(OH)_2]_i$ = 11 mM), the calcium concentration doubled. At the highest C/S, calcium concentration increases from 13 mM to 20 mM at 176 mM of gluconate.

The qualitative evolution of the silicate concentration upon addition of gluconate is similar to that of calcium for all C/S studied. At the lowest C/S ratio (0.75), a moderate increase of silicate concentration from 1.1 mM in the absence of gluconate to 3.2 mM in the presence of 192 mM gluconate is observed. At higher C/S (1.1 and 1.3), the relative increase in silicate concentration in response to gluconate is stronger. The equilibrium silicate concentration with C-S-H of C/S ratio 1.3 in the presence of ~200 mM of gluconate is observed to be two orders of magnitude larger than in the absence of gluconate.

In Figure 8, the calculated Ca and Si concentrations using the Ca-gluconate complexes from

(3) as summarized in Table 1 together with the polynuclear Si-complex, $Ca_3Gluc_2(H_3SiO_4)_2(OH)_2^0$, were used to describe the increase in the experimental calcium and silicate concentrations at equilibrium with C-S-H as a function of gluconate concentration. At all C/S studied, the thermodynamic modeling describes globally well the increase in the Si and Ca concentrations.

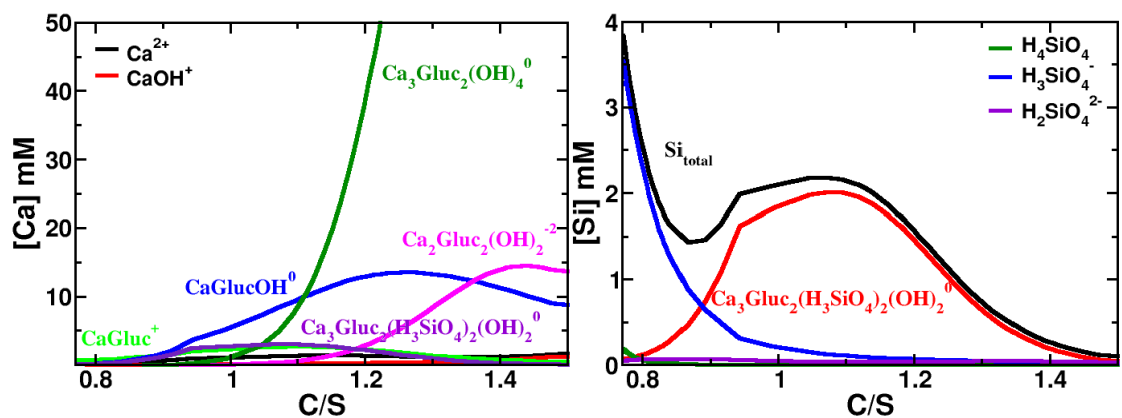

Figure 9: Simulated speciation of calcium and silicate species in a 250 mL solution containing 200 mM of sodium gluconate in equilibrium with 200 mg of C-S-H with varying C/S ratio. Note that both $Ca^{2+}$ and $CaOH^+$ can hardly be seen as there concentration is always lower than 3 mM whatever the C/S. The distribution diagram can be found in the sup. info., see Figure S3 and S4.

Figure 9 highlights the calculated calcium and silicate speciation as a function of the C/S ratio in C-S-H in solutions in equilibrium with 200 mM of sodium gluconate. The calcium concentration is dominated by Ca-gluconate-hydroxide complexes, while the quaternary $Ca_3Gluc_2(H_3SiO_4)_2(OH)_2^0$ complex is present at low concentrations only. In contrast, the experimental and calculated data highlight a visible increase of the silicate complex $Ca_3Gluc_2(H_3SiO_4)_2(OH)_2^0$ in solution, which dominates the silicate speciation at C/S above 0.9, and lower the amount of free silicates species in solution, $H_4SiO_4^0$, $H_3SiO_4^-$, and $H_2SiO_4^-$, in particular at high pH. To further illustrate the importance of the quaternary $Ca_3Gluc_2(H_3SiO_4)_2(OH)_2^0$ complex a comparison of the measured and predicted total equilibrium concentrations of Si and Ca when this complex is ignored is provided in the supplementary materials, see Figure S5.

### 3.2. Adsorption of the organics on C-S-H and portlandite.

Based on the comparison of the measured organic concentrations with the initial concentrations, adsorption isotherms of the gluconate, sorbitol, mannitol and galactitol on C-S-H and $Ca(OH)_2$ (portlandite data are reproduced from (3)) could be obtained as summarized in Figure 10. Gluconate adsorbs strongly on portlandite where an uptake of up 0.3 gluconate per Ca(OH)$_2$ was observed (see Figure 10a), while the sorption of sorbitol, mannitol and galactitol is much weaker. In the case of C-S-H buffered with portlandite (C/S 1.5) an even stronger adsorption of gluconate is observed (up to 0.8 Gluc/Si) and follows a typical Langmuir adsorption. Again the sorption of gluconate is much stronger than the sorption of sorbitol, mannitol and galactitol and reaches up to 0.2 Hex/Si. At low C/S a weaker adsorption is observed for all organics studied in agreement with the decrease in gluconate adsorption on C-S-H with lower C/S reported in literature. (30) (22) (31)

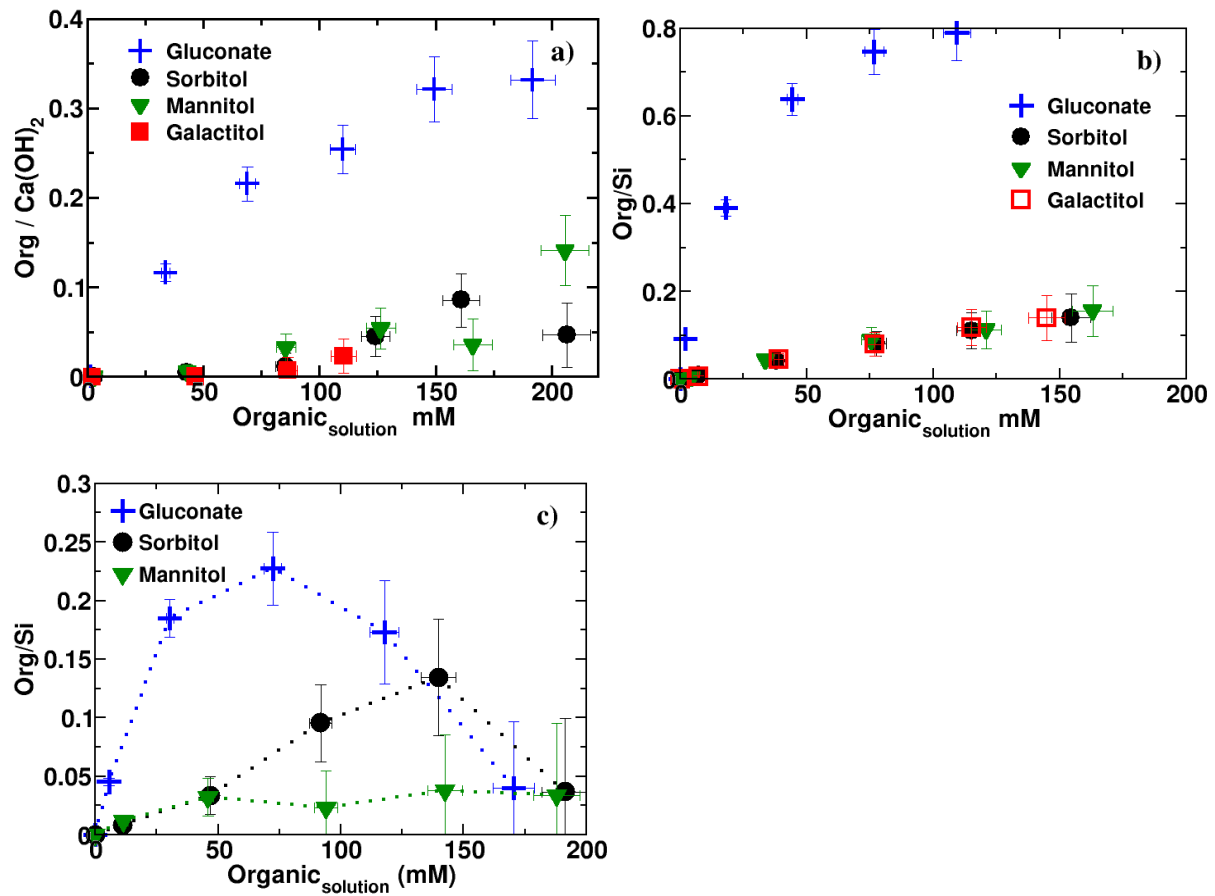

Figure 10: Adsorption isotherms of the organics on: a) portlandite, b) C-S-H buffered with an excess of portlandite and c) C-S-H with an initial C/S = 1.12. For the C-S-H systems, the organic adsorption is expressed as the number ratio of adsorbed organics to C-S-H silicon.

The decrease of gluconate and hexitol sorption with the bulk calcium concentration (i.e. with C/S of C-S-H) clearly indicates that the binding of gluconate and hexitols on C-S-H is mediated by calcium ions, in agreement with the finding of Androniuk et al. (31) on gluconate and of Labbez et al (32)(33) on the adsorption mechanism of anionic polyelectrolyte and sulfate ions. The calcium mediated adsorption is obvious in the case of gluconate since both the latter and C-S-H are negatively charged and even more so as the pH increases (35). As a consequence, adsorption of gluconate (or SO$_4^{2-}$) on C-S-H becomes significant only in conditions where the accumulation of calcium ions on the C-S-H surface is high enough to overcompensate the bare charge of C-S-H. That is at high enough pH and bulk calcium concentration, i.e. above 2 mM of Ca(OH)$_2$ or C/S > 0.95. In the case of the hydrophilic and neutral hexitol molecules, the fact that their binding on C-S-H is most certainly also mediated by calcium ions might seem more surprising at first sight. As we have shown, however, these neutral molecules also form aqueous complexes with calcium ions and their adsorption increases with their complexation strength with calcium. Furthermore, the binding of hexitols on C-S-H is also observed to increase with the calcium content of the solution in the same qualitative manner as that of gluconate (and SO$_4^{2-}$). Finally, the surface affinity of the molecules is found to be correlated to their effectiveness to bind calcium ions in solution. This generic behavior was predicted by Monte Carlo simulations in the case of calcium mediated adsorption of anionic molecules on negatively charged surfaces [32][33][35]. It was recently confirmed in the recent experimental work of Nalet and co-workers whom observed a clear correlation between the calcium binding effectiveness (calcium complexation) and adsorbed amount on C-S-H particles of a large set of anionic molecules [22].

Interestingly also, the adsorption isotherm of organics at C/S = 1.1 is not observed to increase monotonically with the organics concentration as it does in buffered portlandite conditions (C/S = 1.5). Upon progressive addition of organics, it first increases to a maximum and decreases at higher organic additions. Such a behavior is not observed when C-S-H is buffered with portlandite, that is when the calcium hydroxide activity is maintained constant. Instead, when the solution is buffered with portlandite ( constant calcium hydroxide activity) a typical Langmuir isotherm is recovered. This generic behavior was well described and predicted by Turesson and co-authors (32) studying the adsorption of polyanions on negatively charged surfaces in presence of calcium ions with the help of Monte-Carlo simulations in the framework of the primitive model. It can be explained as follows. At C/S = 1.1, there is no calcium reservoir (no portlandite excess) and the calcium (and hydroxide) activity decreases progressively with increasing the organic concentration, because of the increasing formation of calcium-organics aqueous complexes. This complex formation leads to the progressive desorption of the calcium ion from the C-S-H surface, up to the point where the calcium mediated organic adsorption reaches a maximum and finally drops. In presence of an excess of portlandite, the complex formation upon organics addition leads instead to the progressive dissolution of the portlandite, such as to maintain a constant mean activity of calcium hydroxide in the solution. As a result, the adsorption of both calcium and organics increases continuously until the saturation of the C-S-H surface is reached.

## 4. Conclusions

The solubility of C-S-H in the presence of three hexitols (D-sorbitol, D-mannitol, D-galactitol) and sodium gluconate has been measured. The experimental solubility data could be modeled with a speciation model for the liquid phase, while C-S-H was described based on the surface complexation model of Haas and Nonat (23) for the solid phase using the thermodynamic modelling package PHREEQC.

Sorbitol, mannitol, galactitol and gluconate did not only form aqueous complexes with calcium ions, but also polynuclear complexes with silicates, with the exception of mannitol. These polynuclear complexes resulted in a significant increase in the silicate concentration of in equilibrium with C-S-H. The increase in silicate concentration in response to organics addition was found to strengthen with the $Ca(OH)_2$ content in the solution, i.e. with the stoichiometric calcium to silicon ratio (C/S) of C-S-H. In the case of gluconate and sorbitol an increase of up to two orders of magnitude could be observed at the highest organic concentration and C/S ratio studied. With the exception of mannitol, the complexation power of the organic molecules with silicate aqueous species was observed to follow the same order as with calcium ions. That is gluconate > sorbitol >> galactitol > mannitol for the silicate complexation as compared to gluconate >> sorbitol > mannitol > galactitol for the calcium complexation. Interestingly, the complexation strength of the organics with silicates seems to be correlated to their observed retarding effect on $C_3S$ hydration: gluconate >> sorbitol > galactitol > mannitol. (19), but not with complexation strength of the organics with calcium. The fact that both high pH and calcium concentration is required for the complexation of silicate by the organics was well captured with the formation of hetero polynuclear complex involving Ca and OH ions, namely $Ca_2Hex_2(H_3SiO_4)_2(OH)_2^0$, $Ca_2Hex_2(H_2SiO_4)(OH)_4^{-2}$ and $Ca_3Gluc_2(H_3SiO_4)_2(OH)_2$. These Si-complexes when combined with the organic-Ca complexes introduced in our previous work allowed us to describe very satisfactorily the solubility of C-S-H in the presence of the organics.

The adsorption of the organics on C-S-H follows the same trend as on portlandite. The organics affinity with C-S-H was found with the order gluconate >> sorbitol > mannitol ~ galactitol. A typical Langmuir isotherm was found when C-S-H is buffered with $Ca(OH)_2$. The organics adsorption was observed to drop with decreasing the C/S of C-S-H and at high organic concentrations, strongly suggesting a calcium mediated adsorption.

**Acknowledgements**

We would like to thank for the financial support from Nanocem (core project 15). We are also very grateful for many helpful discussions with the representatives of the industrial partners L. Pegado, J.H. Cheung, V. Kocaba, P. Juilland and M. Mosquet and their interest. A special thanks to L. Brunetti, S. El Housseini, K. Alloncle and D. Nguyen for their help in the laboratory work.